\documentclass[12pt]{article}

\usepackage{epsfig}
\input epsf

\title{Higgs boson production at hadron colliders \\in the $k_T$-factorization approach }

\author{A.V.~Lipatov, N.P.~Zotov}

\begin{document}

\maketitle

\begin{center}

{\it D.V.~Skobeltsyn Institute of Nuclear Physics,\\ 
M.V. Lomonosov Moscow State University,
\\119992 Moscow, Russia\/}\\[3mm]

\end{center}

\vspace{1cm}

\begin{center}

{\bf Abstract }

\end{center}

We consider the Higgs boson production at high energy hadron colliders
in the framework of the $k_T$-factorization approach.
The attention is focused on the dominant gluon-gluon fusion subprocess.
We calculate the total cross section and transverse momentum distributions 
of the inclusive Higgs production using unintegrated gluon distributions 
in a proton obtained from the full CCFM evolution equation. 
We show that $k_T$-factorization gives a possibility to 
investigate the associated Higgs boson and jets production.
We calculate the transverse momentum 
distributions and study the Higgs-jet and jet-jet azimuthal 
correlations in the Higgs + one or two jet production processes.
We demonstrate the importance of the higher-order corrections within the 
$k_T$-factorization approach. These corrections should be developed and 
taken into account in the future applications.

\vspace{1cm}

\section{Introduction} \indent 

It is well known that the electroweak symmetry breaking in the Standard 
Model of elementary particle interactions is achieved via the Higgs mechanism. 
In the minimal model there are
a single complex Higgs doublet, where the Higgs boson $H$ is the physical neutral
Higgs scalar which is the only remaining part of this doublet after spontaneous
symmetry breaking. In non-minimal models there are additional charged and neutral 
scalar Higgs particles. The search for the Higgs boson takes important 
part at the
Fermilab Tevatron experiments and will be one of the main fields of 
study at the CERN LHC collider~[1].
The experimental detection of the $H$ will be great 
triumph of the Standard Model of electroweak interactions and will mark 
new stage in high energy physics. 

At LHC conditions, the gluon-gluon fusion $gg \to H$ is the dominant 
inclusive Higgs production mechanism~[2, 3]. In this process, 
the Higgs production
occurs via triangle heavy (top) quark loop. The gluon fusion and weak boson 
fusion ($qq \to qqH$ subprocess via $t$-channel exchange of a 
$W$ or $Z$ bosons) are also expected to be the dominant sources of semi-inclusive
Higgs production (in association with one or two hadronic jets)~[4].
The detailed theoretical studies of such processes are necessary, in particular, to
determine an optimal set of cuts on the final state particles.

It is obvious that the gluon-gluon fusion contribution to the Higgs production
at LHC is strongly dependend on the gluon density $xG(x,\mu^2)$ in a 
proton.
Usually gluon density are described by the Dokshitzer-Gribov-Lipatov-Altarelli-Parizi (DGLAP)
evolution equation~[5] where large logarithmic terms proportional to $\ln \mu^2$ are taken into 
account. The cross sections can be rewritten in terms of hard matrix elements convoluted 
with gluon density functions. In this way the dominant contributions come from diagrams 
where the parton emissions in the initial state are strongly ordered in virtuality. 
This is called collinear factorization, as the strong ordering means that the 
virtuality of the parton entering the hard scattering matrix elements can be neglected 
compared to the large scale $\mu^2$. However, at the LHC energies, typical 
values of the incident gluon momentum fractions $x \sim m_H/\sqrt s \sim 0.008$ (for Higgs boson mass 
$m_H = 120$ GeV) are small, and another large logarithmic terms proportional 
to $\ln 1/x$ become important. 
These contributions can be taken into account using Balitsky-Fadin-Kuraev-Lipatov (BFKL) 
evolution equation~[6].
Just as for DGLAP, in this way it is possible to factorize
an observable into a convolution of process-dependent hard matrix elements
with universal gluon distributions. But as the virtualities (and transverse 
momenta) of the propagating gluons are no longer ordered, the matrix 
elements have to be taken off-shell and the convolution made also over 
transverse momentum ${\mathbf k}_T$ with the unintegrated ($k_T$-dependent) gluon 
distribution ${\cal F}(x,{\mathbf k}_T^2)$. The unintegrated gluon distribution
${\cal F}(x,{\mathbf k}_T^2)$ determines the probability to find a gluon carrying the 
longitudinal momentum fraction $x$ and the transverse momentum ${\mathbf k}_T$. 
This generalized factorization is called $k_T$-factorization~[7--10]. 
It is expected that BFKL evolution gives the theoretically correct description at 
assymptotically large energies (i.e. very small $x$). 
At the same time another approach, valid for 
both small and large $x$, have been developed by Ciafaloni, Catani, Fiorani and Marchesini,
and is known as the CCFM model~[11]. It introduces angular ordering of emissions
to correctly treat gluon coherence effects. In the limit of 
asymptotic energies it is almost equivalent to 
BFKL~[12--14], but also similar to
the DGLAP evolution for large $x$ and high $\mu^2$. The resulting unintegrated
gluon distribution depends on two scales, the additional scale ${\bar q}^2$
is a variable related to the maximum angle allowed
in the emission and plays the role of the evolution scale $\mu^2$ in the
collinear parton densities. The following classification
scheme~[15] is used: ${\cal F}(x,{\mathbf k}_T^2)$ denote pure BFKL-type unintegrated 
gluon distributions and ${\cal A}(x,{\mathbf k}_T^2,\mu^2)$ stands for any other 
type having two scale involved. In this paper we will apply the CCFM gluon evolution
to study of the inclusive and semi-inclusive Higgs production at LHC conditions. 

In the collinear factorization,
the calculation of such processes is quite complicated even at lowest order 
because of the heavy quark loops contribution. For example, in Higgs + one jet production, 
triangle and box loops
 occur, and in Higgs + two jet production the pentagon loops occur~[16]. However, 
the calculations of the Higgs production rates can be simplified in the limit of large
top quark mass $m_t \to \infty$~[17]. In this approximation the coupling of the 
gluons to the Higgs
via top-quark loop can be replaced by an effective coupling. Thus it reduces
the number of loops in a given diagram by one. The large $m_t$ approximation is 
valid to an accuracy of $\sim 5$\% in the intermediate Higgs mass range $m_H < 2 m_t$,
as long as transverse momenta of the Higgs or final jets are smaller than of 
the top quark mass ($p_T < m_t$)~[16]. Within this approach, the total cross section 
for $gg \to H + X$ is known to next-to-next-to-leading order (NNLO) accuracy~[18].
Higher-order QCD corrections to inclusive Higgs production were found to be large: their 
effect increases the leading order cross section by about $80 - 100$\%~[19] (see also~[20]). 

A particularly interesting quantity is the transverse momentum 
distribution of the produced Higgs boson. The precise theoretical prediction of the 
$d\sigma/d p_T$ at the LHC is important for quantitative evaluation of the 
required measurement accuracies and detector performance.
It is well-known that the fixed-order perturbative QCD is applicable
when the Higgs transverse momentum is comparable to the $m_H$. Hovewer, the main part of the
events is expected in the small-$p_T$ region ($p_T \ll m_H$), where the coefficients
of the perturbative series in $\alpha_s$ are enhanced by powers of large 
logarithmic terms proportional to $\ln m_H^2/p_T^2$. Therefore reliable predictions at small $p_T$
can only be obtained if these terms will be resummed to all orders. Such procedure 
is called soft-gluon resummation~[21-23] and has been performed in 
collinear calculations at leading logarithmic (LL), next-to-leading logarithmic (NLL)~[24] and 
next-to-next-to-leading logarithmic (NNLL)~[25] levels. Recently it was 
shown~[26] that in the framework of $k_T$-factorization approach the soft 
gluon resummation formulas are the result of the approximate treatment of 
the solutions of the CCFM  evolution equation (in the $b$-representation). 

There are several additional motivations for our study of
the Higgs production in the $k_T$-factorization approach. First of all, in 
the standard collinear approach, when the transverse momentum of 
the initial gluons is neglected, the transerse momentum of the final Higgs
boson in $gg \to H$ subprocess is zero. Therefore it is necessary to 
include an initial-state QCD radiation to generate the $p_T$ 
distributions. It is well known at present that  the 
$k_T$-factorization 
naturally includes a large part of the high-order  perturbative QCD  
corrections~[27].
This fact is illustrated more detailed in Figure~1, which is a schematical
representation of a typical Higgs + jet production process. Figure~1 (a)  shows 
the 
fixed-order perturbative QCD picture where the upper part of the diagram (above 
the dash-dotted line) corresponds to the $gg \to gH$ subrocess, and the lower
part describes the gluon evolution in a proton. As the incoming gluons are assumed
to have zero transverse momentum, the transverse momentum distributions of the 
produced Higgs and jet are totally determined by the properties of the ${\cal O}(\alpha_s^3)$ 
matrix element. In the $k_T$-factorization approach (Figure~1 (b)), the underlying 
partonic subprocess is $gg \to H$, which is formally of order ${\cal O}(\alpha_s^2)$.
Some extra powers of $\alpha_s$ are hidden in the gluon evolution represented by
the part of the diagram shown below the dash-dotted line. In contrast with the collinear
approximation, the $k_T$-factorization takes into account the gluon transverse motion.
Since the upper gluon in the parton ladder is not included in the hard interaction, its
transverse momentum is now determined by the properties of the evolution equation only.
It means that in the $k_T$-factorization approach the study  of transverse 
momenta distributions in the Higgs production
via gluon-gluon fusion will be direct probe of the unintegrated gluon distributions 
in a proton. In this case 
the transverse momentum of the produced Higgs should be equal to the sum of the transverse momenta 
of the initial gluons. Therefore future experimental studies at LHC can be used 
as further test of the non-collinear parton evolution.

In the previous studies~[26, 28, 29] the $k_T$-factorization formalism was 
applied to calculate transverse momentum distribution of the inclusive Higgs production. 
The simplified solution of the CCFM equation in single loop approximation~[30] (when 
small-$x$ effects can be neglected) were used in ~[26]. In such approximation the CCFM evolution 
is reduced to the DGLAP one with the difference that the single loop evolution takes the gluon 
transverse momentum $k_T$ into account. Another simplified solution of the CCFM equation 
was proposed in Ref.~[31], where the transverse momenta of the incoming 
gluons are generated in the last evolution step (Kimber-Martin-Ryskin prescription).
The calculations~[26, 29] were done using the on-mass shell (independent 
from the gluon $k_T$) matrix element of the $gg\to H$ subprocess and rather the 
similar results have been obtained. In Ref.~[28] in the framework of MC generator 
CASCADE~[32] the off-mass-shell matrix element obtained by F.~Hautmann~[33] has
been used with full CCFM evolution. 

In present paper we investigate Higgs production at hadron colliders 
using the full CCFM-evolved unintegrated gluon densities~[28].
We obtain the obvious expression for the $g^* g^* \to H$ off-mass-shell matrix 
element in the large $m_t$ limit apart from Ref.~[33]. After that, we 
calculate the total cross section and transverse momentum 
distribution of the inclusive Higgs production at Tevatron and LHC.
To illustrate the fact that in the $k_T$-factorization approach
the main features of collinear higher-order pQCD corrections are taken into 
account effectively, we give theoretical 
predictions for the Higgs + one jet and Higgs + two jet production processes 
using  some physically motivated approximation.

In Section 2 we recall the basic
formulas of the $k_T$-factorization formalism with a brief review of
calculation steps. In Section 3 we present the numerical results of
our calculations and discussion.
Finally, in Section 4, we give summary of our results.

\section{Basic formulas} \indent 

We start from the effective Lagrangian for the Higgs boson coupling to 
gluons~[16]:
$$
  {\cal L}_{\rm {eff}} = {\alpha_s \over 12 \pi}\left(G_F \sqrt 2\right)^{1/2} G_{\mu \nu}^a G^{a\,\mu \nu} H, \eqno (1)
$$

\noindent 
where $G_F$ is the Fermi coupling constant, $G_{\mu \nu}^a$ is the gluon field strength tensor 
and $H$ is the Higgs field. The triangle vertex $T^{\mu \nu}(k_1,k_2)$ for two off-shell gluons having four-momenta
$k_1$ and $k_2$ and color indexes $a$ and $b$ respectively, can be  
obtained  easily from the Lagrangian (1):
$$
  T^{\mu \nu}(k_1,k_2) = i \delta^{a b} {\alpha_s \over 3\pi} \left(G_F \sqrt 2\right)^{1/2} \left[ k_2^{\mu} k_1^{\nu} - 
    (k_1 \cdot k_2) g^{\mu \nu} \right]. \eqno (2)
$$

\noindent 
To calculate the squared off-mass-shell matrix element for the $g^* g^* \to H$ 
subprocess 
it is necessary to take into account the non-zero virtualities of the initial gluons 
$k_1^2 = - {\mathbf k}_{1T}^2 \neq 0$, $k_2^2 = - {\mathbf k}_{2T}^2 \neq 0$. 
We have obtained\footnote{We would like to remark that the expression (3) 
differs from the one obtained in Ref.~[33].}  
$$
  |\bar {\cal M}|^2(g^* g^* \to H) = {\alpha_s^2(\mu^2)\over 576 \pi^2} G_F \sqrt 2  
    \left[ m_H^2 + {\mathbf k}_{1T}^2 + {\mathbf k}_{2T}^2 + 2 |{\mathbf k}_{1T}||{\mathbf k}_{2T}|\cos \phi \right]^2 \cos^2 \phi, \eqno (3)
$$

\noindent 
where $\phi$ is the azimuthal angle between transverse momenta ${\mathbf k}_{1T}$ and
${\mathbf k}_{2T}$, the transverse momentum of the produced Higgs boson is
${\mathbf p}_{T} = {\mathbf k}_{1T} + {\mathbf k}_{2T}$ and the virtual gluon 
polarization tensor has been taken in the form~[7, 8] 
$$
  \sum \epsilon^{\mu} \epsilon^{*\,\nu} = {k_T^{\mu} k_T^{\nu} \over {\mathbf k}_T^2 }. \eqno (4)
$$

\noindent
The cross section of the inclusive Higgs production $p\bar p \to H + X$ in the $k_T$-factorization 
approach can be written as
$$
  \displaystyle d\sigma(p \bar p \to H + X) = \int {dx_1\over x_1} {\cal A}(x_1,{\mathbf k}_{1T}^2,\mu^2) d{\mathbf k}_{1T}^2 {d\phi_1 \over 2\pi} \times \atop 
    \displaystyle \times \int {dx_2\over x_2} {\cal A}(x_2,{\mathbf k}_{2T}^2,\mu^2) d{\mathbf k}_{2T}^2 {d\phi_2 \over 2\pi} d\hat \sigma(g^* g^* \to H), \eqno (5)
$$

\noindent
where $\hat \sigma(g^* g^* \to H)$ is the Higgs production
cross section with off-mass-shell gluons, $x_1$ and $x_2$ are the 
longitudinal momentum fractions, 
and ${\cal A}(x,{\mathbf k}_{T}^2,\mu^2)$ is the
unintegrated gluon distributions in a proton. Let $s = (p_1 + p_2)^2$ and $p_1$, 
$p_2$ 
are  the four-vectors of the incoming protons. Then the  differential
cross section reads
$$
  \displaystyle {d\sigma(p \bar p \to H + X)\over dy_H} = \int {\alpha_s^2(\mu^2)\over 288 \pi} {G_F \sqrt 2 \over x_1 x_2 m_H^2 s} \left[m_H^2 + {\mathbf p}_T^2\right]^2 \cos^2 \phi_2 \times \atop
    \displaystyle \times {\cal A}(x_1,{\mathbf k}_{1T}^2,\mu^2) {\cal A}(x_2,{\mathbf k}_{2T}^2,\mu^2) d{\mathbf k}_{1T}^2 d{\mathbf k}_{2T}^2 {d\phi_2 \over 2\pi}, \eqno (6) 
$$

\noindent
where $y_H$ is the Higgs rapidity in the proton-proton c.m. frame. The longitudinal momentum
fractions $x_1$ and $x_2$ are given by
$$
  x_1 = \sqrt{m_H^2 + {\mathbf p}_T^2\over s} \exp (y_H),\quad x_2 = \sqrt{m_H^2 + {\mathbf p}_T^2\over s} \exp (-y_H). \eqno (7) 
$$

\noindent
If we average the expression (6) over transverse momenta ${\mathbf k}_{1T}$ and 
${\mathbf k}_{2T}$ and take the limit ${\mathbf k}_{1T}^2 \to 0$, ${\mathbf k}_{2T}^2 \to 0$,
we obtain well-established expression~[2] for Higgs production cross section in leading-order 
perturbative QCD:
$$
  d\sigma(p \bar p \to H + X) = {\alpha_s^2(\mu^2)\over 576 \pi} G_F \sqrt 2 {m_H^2 \over x_1 x_2 s} x_1{\cal G}(x_1,\mu^2) x_2{\cal G}(x_2,\mu^2) dy_H, \eqno (8) 
$$

\noindent
where $x {\cal G}(x,\mu^2)$ is the usual
(collinear) gluon density which is related with the unintegrated gluon 
distribution 
${\cal A}(x,{\mathbf k}_{T}^2,\mu^2)$  by
$$
  x {\cal G}(x,\mu^2) \sim \int {\cal A}(x,{\mathbf k}_{T}^2,\mu^2) d{\mathbf k}_{T}^2. \eqno (9)
$$

\noindent
Here the sign $\sim$ indicates, that there is no strict
equality between the left and the right parts of the equation (9)\footnote{See 
Refs.~[15, 34] for more details.}.

The multidimensional integration in the expression (6) has been performed
by means of the Monte Carlo technique, using the routine VEGAS~[35].
The full C$++$ code is available from the authors on
 request\footnote{lipatov@theory.sinp.msu.ru}.

\section{Numerical results and discussion} 

\subsection{Inclusive Higgs production} \indent

We now are in a position to present our numerical results. First we
describe our theoretical input and the kinematical conditions.               
Besides the Higgs mass $m_H$, the cross section (6) depend on the uninterated gluon 
distribution ${\cal A}(x,{\mathbf k}_{T}^2,\mu^2)$ and the energy scale $\mu$.
The new fits of the unintegrated gluon density (J2003 set 1 --- 3) have
 been recently presented~[28].
The full CCFM equation in a proton was solved numerically using a Monte Carlo method.
The input parameters were fitted to describe the proton structure function
$F_2(x,Q^2)$. Since these gluon densities reproduce well
 the forward jet production at HERA, charm
 and bottom production data at
Tevatron~[28] and charm and $J/\psi$ production at LEP2 energies~[35], we 
use it (namely J2003 set 1) in our calculations.
As is often done for Higgs production, we choose the 
renormalization and factorization scales to be $\mu = \xi m_H$, and
vary the scale parameter $\xi$ between $1/2$ and $2$ about the default value $\xi = 1$.
Also we use LO formula for the 
strong coupling constant $\alpha_s(\mu^2)$ with $n_f = 4$ active quark flavours 
and $\Lambda_{\rm QCD} = 200$ MeV, such that $\alpha_s(M_Z^2) = 0.1232$.

In Figure~2 and~3 we display our prediction for the transverse momentum and rapidity distributions
of the inclusive Higgs production at the LHC ($\sqrt s = 14$ TeV).
The calculations were done for four choices of the Higgs boson mass under
interest in the Standard Model with default scale $\mu^2 = m_H^2$. 
The solid, dashed, dash-dotted and dotted lines
correspond $m_H = 125$ GeV, $m_H = 100$ GeV, $m_H = 150$ GeV (where $WW$ decay 
channel is dominant) and $m_H = 200$ GeV (above $WW$ and $ZZ$ decay tresholds), 
respectively. One can see that mass effects are present only at low $p_T < m_H$, whereas
all curves practically coincide at large transverse momenta.
We note that our predictions which correspond to the Higgs mass $m_H = 125$ GeV slightly underestimate results obtained in the 
combined fixed-order + resummed approach~[37].
In this approach
fixed-order predictions (at LO or NLO level) and 
resummed ones (at NLL or NNLL level, respectively) have to be consistenly 
matched at moderate $p_T$. The NNLL + NLO results~[25] are smaller than
NLL + LO ones~[24] by about $20$\% at low transverse momenta.
We see that our predictions lie below NNLL + NLO calculations by 
about $15$\% in this kinematical region.
Usage the doubly unintegrated gluon distributions results in
 more flat behaviour of the $p_T$-distribution~[29] in comparison with 
both our and NNLL + NLO predictions.

We note also that the peak in the transverse momentum distribution occurs at a smaller value of $p_T$
compared to the NNLL + NLO calculations. The location of this peak as a function
of Higgs boson mass is shown in Figure~4. We find that at $m_H = 125$ GeV the peak occurs
at $p_T \sim 10$ GeV, whereas NNLL + NLO line peaks at $p_T \sim 15$ 
GeV~[37].
The similar effect has been obtained~[29] when doubly unintegrated gluon 
distributions were used.

The total cross sections of the inclusive Higgs production at Tevatron 
($\sqrt s = 1.96$ TeV) and LHC 
conditions as function Higgs mass are plotted in Figure 5 and 6 in the 
mass range $m_H = 100 - 200$ GeV.
The solid lines are obtained by fixing both the factorization and renormalization 
scales at the default value $\mu = m_H$. In order to estimate the 
theoretical uncertainties in our predictions, we vary the unphysical parameter
$\xi$ as indicated above. These uncertainties are presented by upper and
lower dashed lines. We find that our default predictions agree very well with 
recent NNLO results~[18]. For example, when Higgs boson mass is $m_H = 
120$ GeV, 
our calculations give $\sigma = 0.84$ pb at Tevatron and $\sigma = 35.9$ pb at LHC.
However, the scale dependences are rather large. At LHC energy, 
it changes from about $20$\% when $m_H = 100$ GeV, to about $50$\% when 
$m_H = 200$ GeV. At Tevatron, it range from $40$\% to $50$\%, respectively.
This fact indicates the
necessarity of high-order corrections inclusion in the
$k_T$-factorization formalism. But  one should note that
in the $k_T$-factorization the role of such correction 
is very different in comparison with the corrections in the collinear 
approach, since
part of the standard high-order corrections are already included
at LO level in $k_T$-factorization\footnote{See also~[15, 34] for more 
detailed discussion.}.
At the same time the theoretical uncertainties of the collinear QCD calculations,
after inclusion of both NNLO corrections and soft-gluon 
resummation at the NNLL level, are about $10$\% in the 
low mass range $m_H < 200$ GeV~[18].

\subsection{Higgs production in association with jets} \indent

Now we demonstrate how $k_T$-factorization approach can be 
used to calculate the semi-inclusive Higgs production rates. The produced Higgs boson
is accompanied by a number of gluons radiated in the course of the gluon evolution.
As it has been noted in~Ref.~[38], on the average the gluon 
transverse 
momentum decreases from the hard interaction block towards the proton. As an
approximation, we assume that the gluon $k'$ closest to the Higgs compensates
the whole transverse momentum of the virtual gluon participating in the
gluon fusion, i.e. ${\mathbf k'}_T \simeq - {\mathbf k}_T$ (see Figure~1).
All the other emitted gluons are collected together in the proton remnant, 
which is assumed to carry only a negligible transverse momentum compared
to ${\mathbf k'}_T$. This gluon gives rise to a final hadron jet with 
${\mathbf p}_{{\rm jet}\,T} = {\mathbf k'}_T$.

From the two hadron jets represented by the gluons from the 
upper and lower evolution ladder we choose the one carrying the largest
transverse momentum, and then compute Higgs with an associated
jet cross sections at the LHC energy. We have 
applied the usual cut on the final jet transverse momentum 
$|{\mathbf p}_{{\rm jet}\,T}| > 20$ GeV. Our predictions for the 
transverse momentum distribution of the Higgs + one jet production are 
shown in Figure 7. 
As in the inclusive Higgs production case, we test four different $m_H$ values
in the transverse momentum ditributions.
All curves here are the same as in Figure~2.
One can see the shift of the peak position in the
$p_T$ distributions in comparison with inclusive production, which is direct 
consequence of the $|{\mathbf p}_{{\rm jet}\,T}| > 20$ GeV cut.
We note that the rapidity interval between the 
jet and the Higgs boson is naturally large. It is because there is 
angular ordering 
in the CCFM evolution, which is equivalent to an ordering in rapidity 
of the emitted gluons.

The investigation of the different azimuthal correlations between 
final particles in semi-inclusive Higgs production provides many interesting 
insights. In particular, studying of these quantities are important to clean
separation of weak-boson fusion and gluon-gluon fusion contributions.
To demonstrate the possibilities of the $k_T$-factorization approach,
we present here the two azimuthal angle distributions. First, we calculate
azimutal angle distribution between the Higgs boson and final jet transverse 
momenta in the Higgs + one jet production process. Second, we
calculate azimuthal angle distributions between the two final jet transverse 
momenta in the Higgs + two jet production process.
In this case the Higgs boson is centrally located in rapidity between the two jets 
and it is very far from either jet, and the kinematical cut $|{\mathbf 
p}_{{\rm jet}\,T}| > 20$ 
GeV was applied for both final jets. We set no cuts on the jet-jet 
invariant mass.
Our results are shown in Figure~8 and 9, respectively. Figure~8 
demonstrated roughly the back-to-back Higgs + one jet production. In 
Figure~9 we obtained a dip at $90$ degrees 
in jet-jet azimuthal correlation, which is characteristic for 
loop-induced Higgs coupling to gluons~[39]. The fixed-order perturbative 
QCD 
calculations of the $gg \to gg H$ subprocess give the similar result~[16].
However, as it was already mentioned above, such calculations
are very cumbersome even at leading order. The evaluation of the 
radiative corrections at ${\cal O}(\alpha_s)$ to Higgs + two jet
production would imply the calculation of up to hexagon quark loops
and two-loop pentagon quark loops, which are at present unfeasible~[20].
We note that contribution from the weak-boson fusion to the Higgs + two jet 
production has flat behavior of the jet-jet angular distribution~[16, 20].

To illuminate the sensitivity of the Higgs production rates 
to the details of the unintegrated gluon distribution, we 
 repeated our calculations for jet-jet angular correlations 
using J2003 set 2 gluon density~[28] (dashed line in Figure~9). This 
density
takes into account the singular and non-singular terms in the CCFM splitting
function, where the Sudakov and non-Sudakov form factors
 were modified accordinly. We note that J2003 set 1 takes 
into account only singular terms.
Both these sets describe the proton structure function $F_2(x,Q^2)$ at HERA reasonable well.
However, one can see the very large discrepancy (about order of magnitude) 
between predictions of J2003 set 1 and set 2 unintegrated gluon densities.
The similar difference was claimed~[28] for charm and bottom production at 
Tevatron also.
This fact clearly indicates again that high-order corrections 
to the leading order $k_T$-factorization are important and should be
developed for future applications.

\section{Conclusions} \indent 

We have considered the Higgs boson production via gluon-gluon fusion 
at high energy hadron colliders in the framework of the 
$k_T$-factorization approach. Our interests were focused on the
Higgs boson total cross section and transverse momenta distributions at 
Tevatron and LHC colliders. 
In our numerical calculations we use the J2003 set 1 unintegrated
gluon distribution, which was obtained recently from the full CCFM
evolution equation.

We find that $k_T$-factorization gives the very close to NNLO pQCD results
for the inclusive Higgs production total cross sections. It is because
the main part of the high-order collinear pQCD corrections is already
included in the $k_T$-factorization. 
Also we have demonstrated that $k_T$-factorization gives a possibility to 
investigate the associated Higgs boson and jets production in much more
simple manner, than it can be done in the collinear factorization.
Using some approximation, we have calculated transverse momentum 
distributions and investigated the Higgs-jet and jet-jet azimuthal 
correlations in the Higgs + one or two jet production processes.
However, the scale 
dependence of our calculations is rather large (of the order of $20 - 50$\%), 
which indicates the importance of the high-order correction within the 
$k_T$-factorization approach. These corrections should be developed and 
taken into account in the future applications.

We point out that in this paper we do not try to give a better prediction
for Higgs production than the fixed-order pQCD calculations.
The main advantage of our approach is that it is possible to
obtain in straighforward manner the analytic description which
reproduces the main features of the collinear high-order pQCD 
calculations\footnote{In this part our conclusions coincide with ones from
Ref.~[29].}.
But in any case, the future experimental study of such
processes at LHC will give important information about non-collinear gluon
evolution dynamics, which will be useful even for leading-order 
$k_T$-factorization formalism.

\section{Acknowledgements} \indent 

The authors are very grateful to H. Jung for possibility to use the CCFM 
code 
for  unintegrated gluon distributions in our calculations, for reading of 
the manuscript and useful discussion. We thank S.P. 
Baranov for encouraging interest and helpful discussions. N.Z. thanks P.F. Ermolov for
support
and the DESY directorate for the hospitality and support.

\newpage

\begin{figure}
\begin{center}
\epsfig{figure=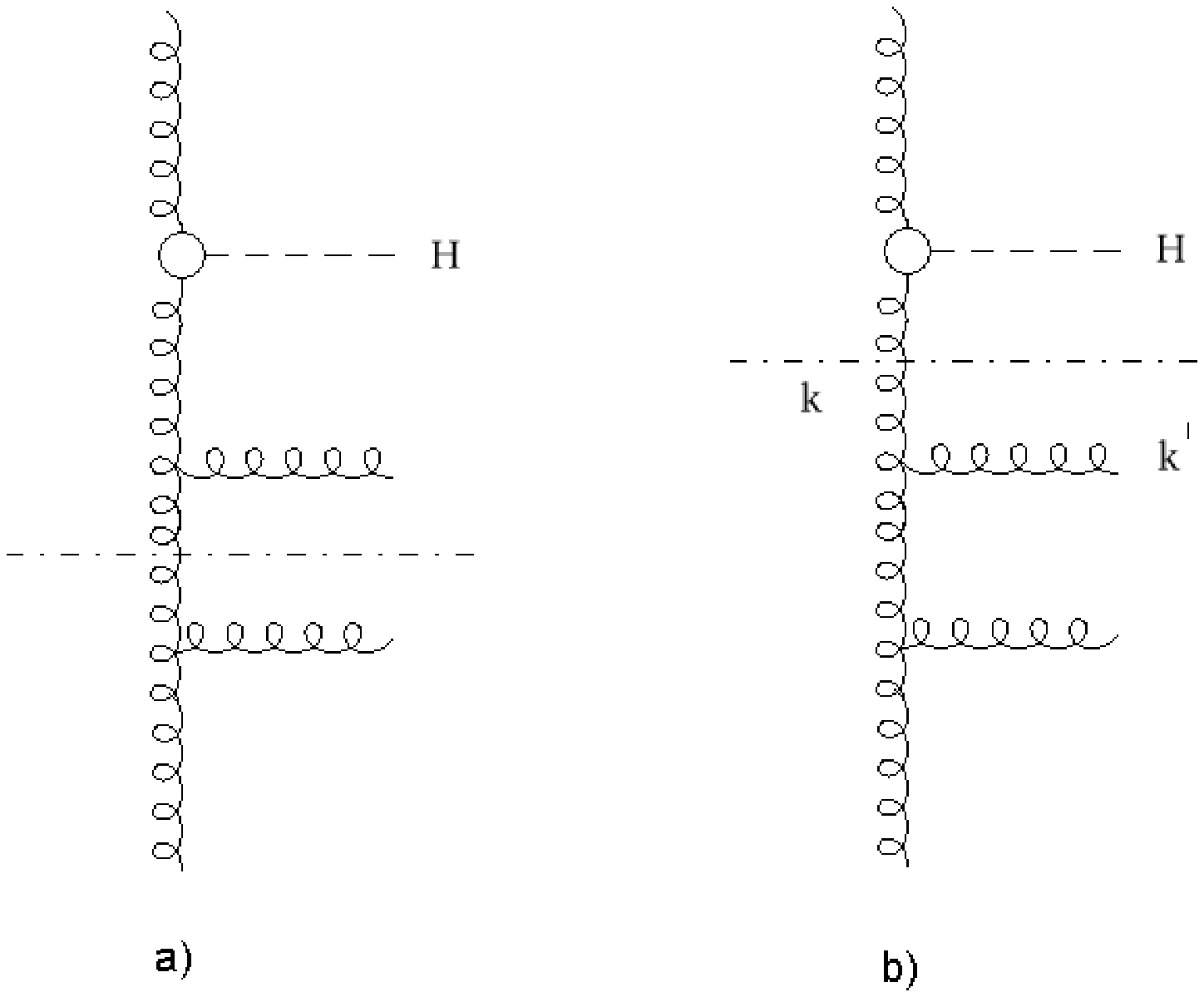}
\end{center}
\caption{The typical Feynman diagram contributing to the Higgs boson 
production in the collinear (a) and $k_T$-factorization (b) approaches.}
\label{fig1}
\end{figure}

\newpage
 
\begin{figure}
\begin{center}
\epsfig{figure=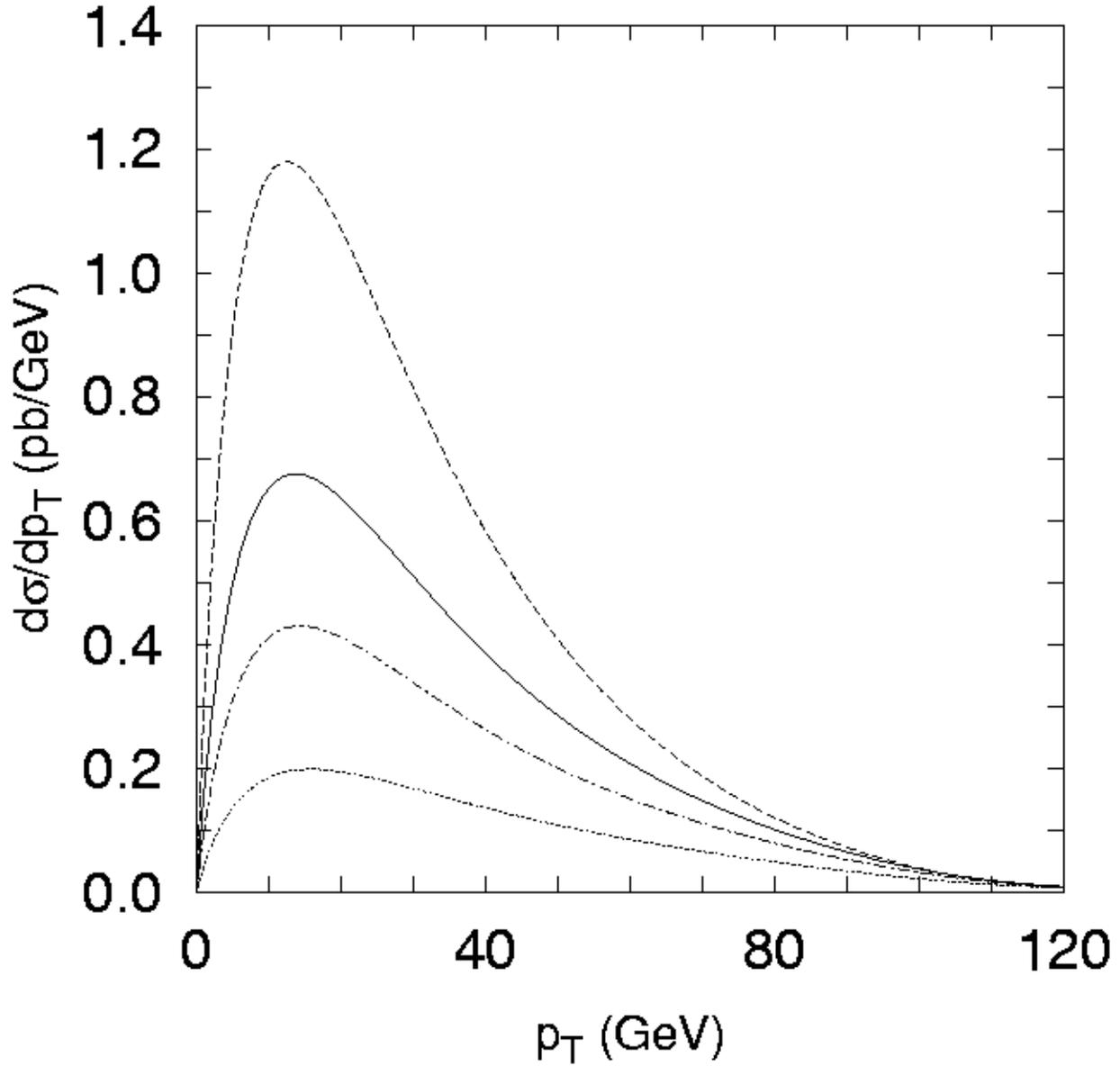}
\end{center}
\caption{Differential cross section $d\sigma/dp_T$ for inclusive Higgs boson production 
  at $\sqrt s = 14$ TeV. The solid, dashed, dash-dotted and dotted lines
  correspond $m_H = 125$ GeV, $m_H = 100$ GeV, $m_H = 150$ GeV and $m_H = 200$ GeV, 
  respectively.}
\label{fig2}
\end{figure}

\newpage

\begin{figure}
\begin{center}
\epsfig{figure=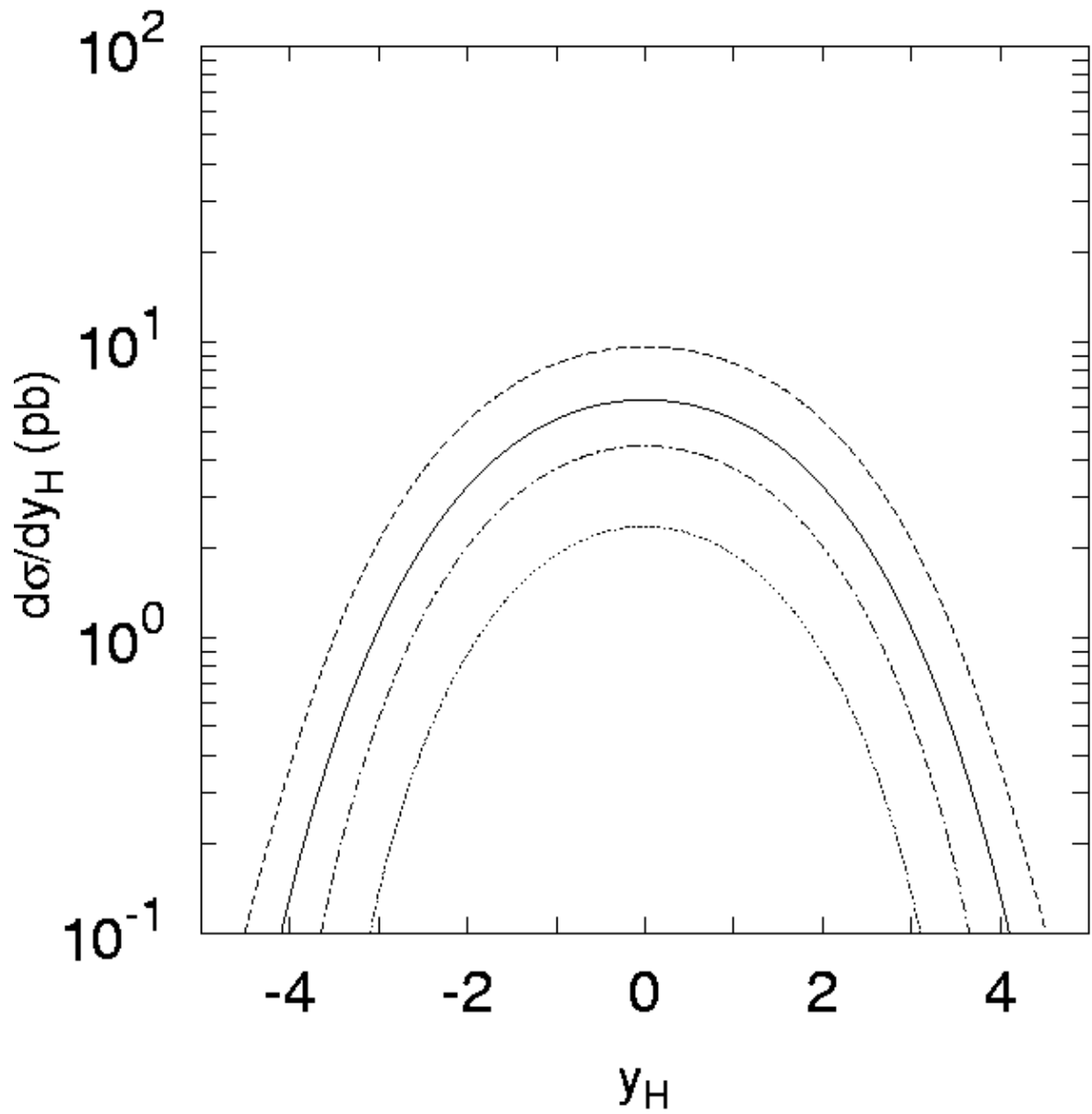}
\end{center}
\caption{Differential cross section $d\sigma/dy_H$ for inclusive Higgs boson production 
  at $\sqrt s = 14$ TeV. All curves are the same as in Figure 2.}
\label{fig3}
\end{figure}

\newpage

\begin{figure}
\begin{center}
\epsfig{figure=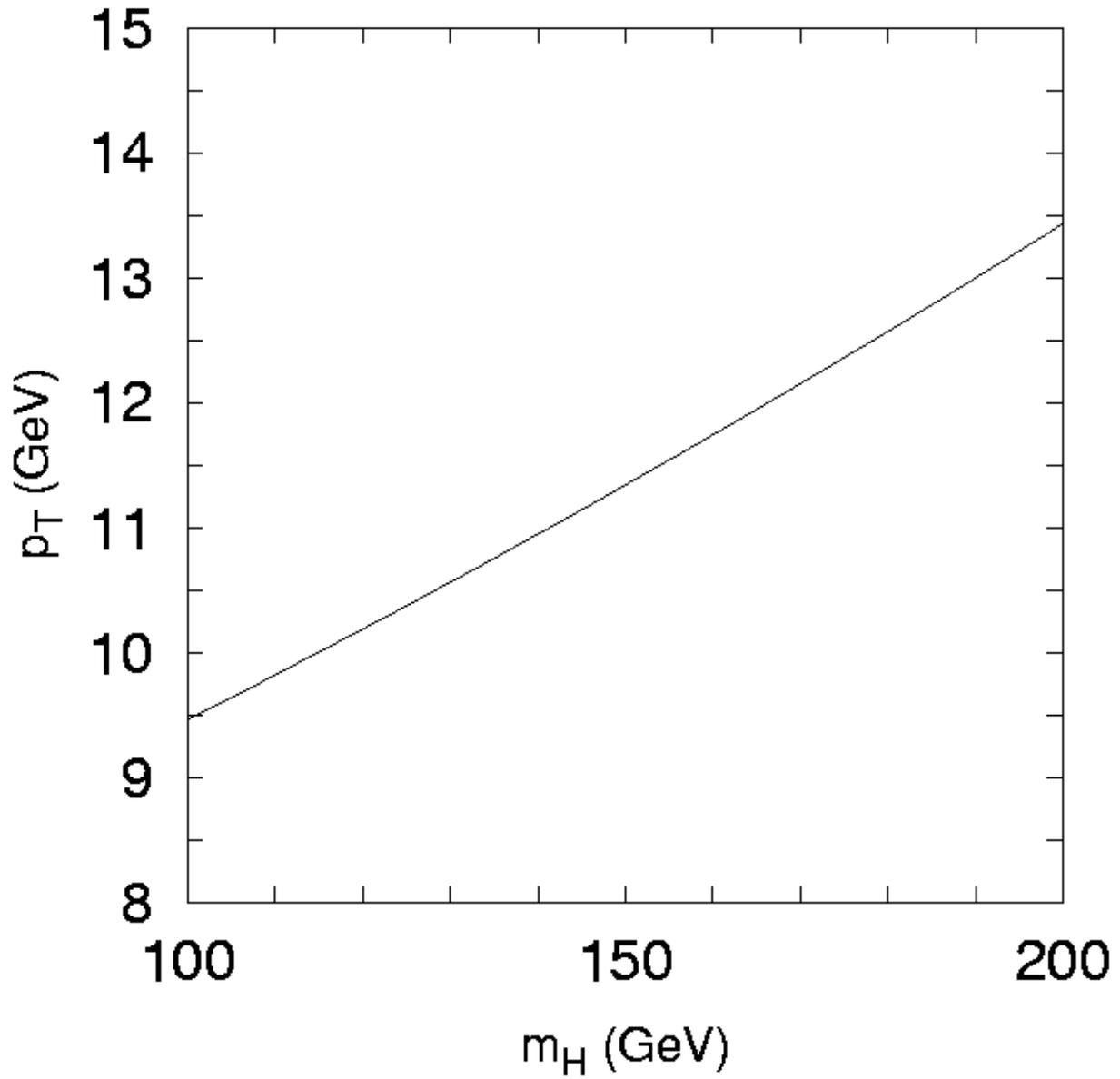}
\end{center}
\caption{Location of the peak of the transverse momentum distributions for inclusive 
  Higgs boson production as a function of Higgs mass at $\sqrt s = 14$ TeV.}
\label{fig4}
\end{figure}

\begin{figure}
\begin{center}
\epsfig{figure=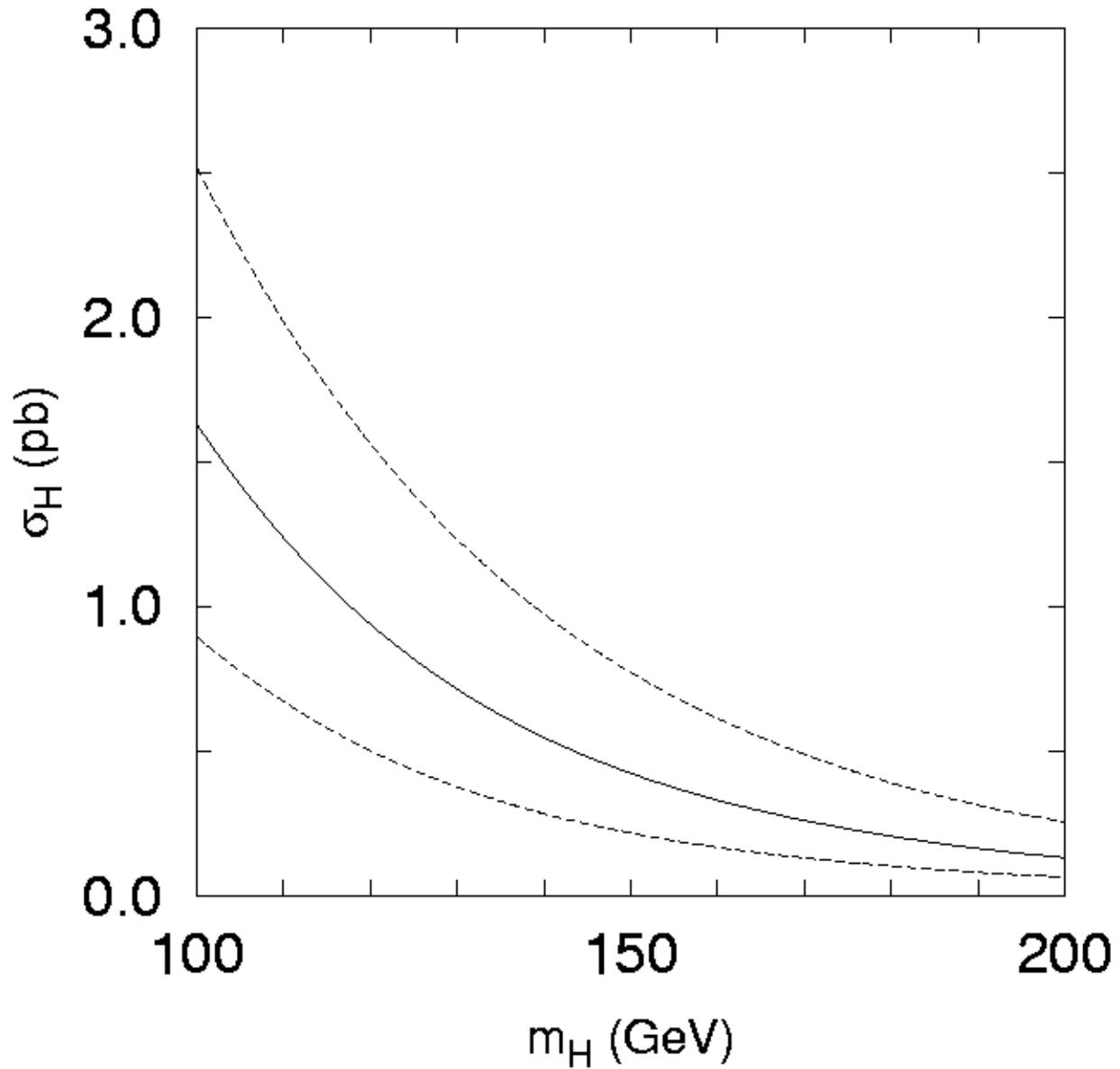}
\end{center}
\caption{Total cross section of inclusive Higgs boson production as a function
  of Higgs mass at $\sqrt s = 1.96$ TeV. The solid line corresponds to the default 
  scale $\mu = m_H$, whereas upper and lower dashed lines correspond to the 
  $\mu = m_H/2$ and $\mu = 2 m_H$ scales, respectively.}
\label{fig5}
\end{figure}

\begin{figure}
\begin{center}
\epsfig{figure=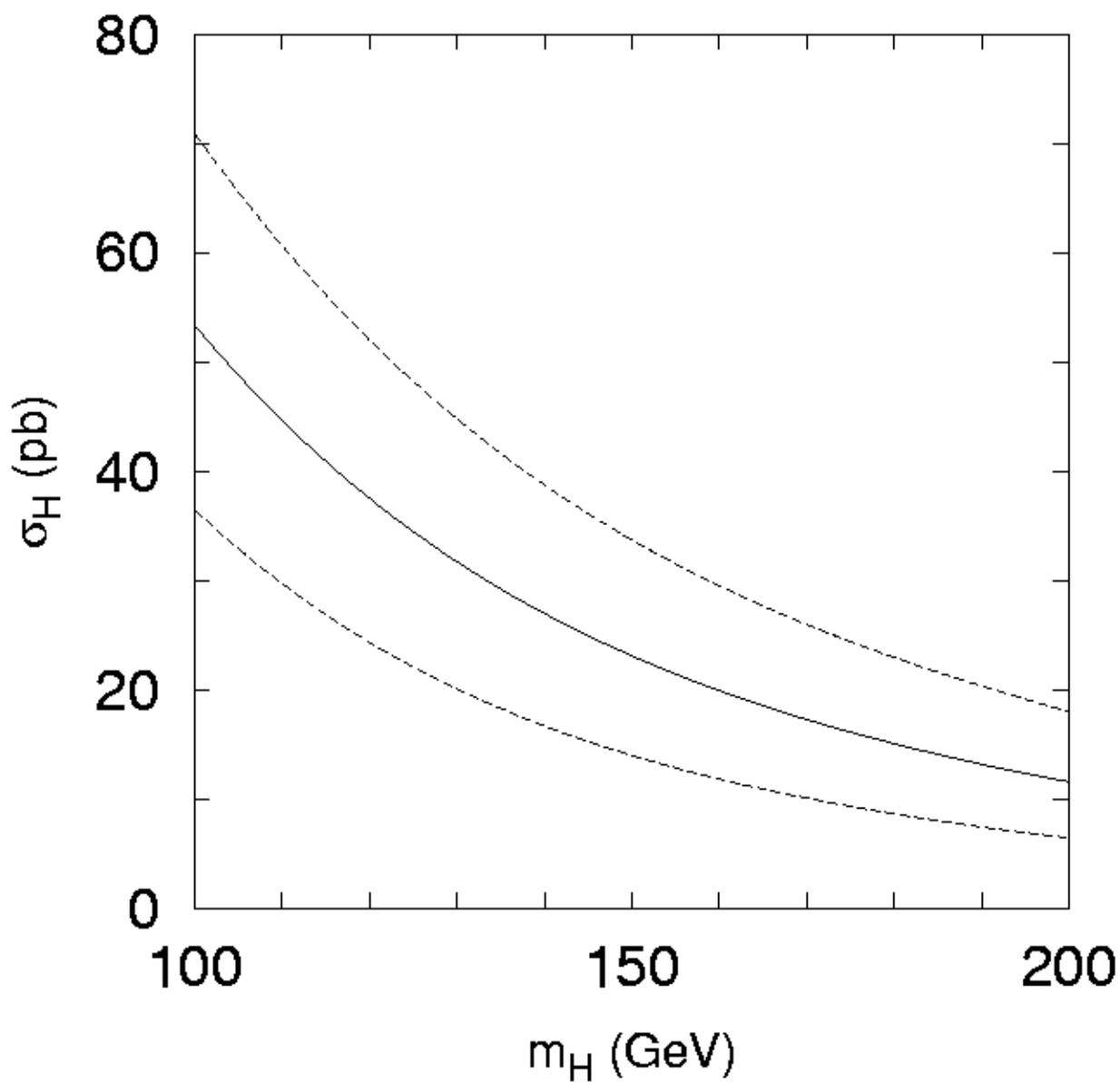}
\end{center}
\caption{Total cross section of inclusive Higgs boson production as a function
  of Higgs mass at $\sqrt s = 14$ TeV. All curves are the same as in Figure 5.}
\label{fig6}
\end{figure}

\begin{figure}
\begin{center}
\epsfig{figure=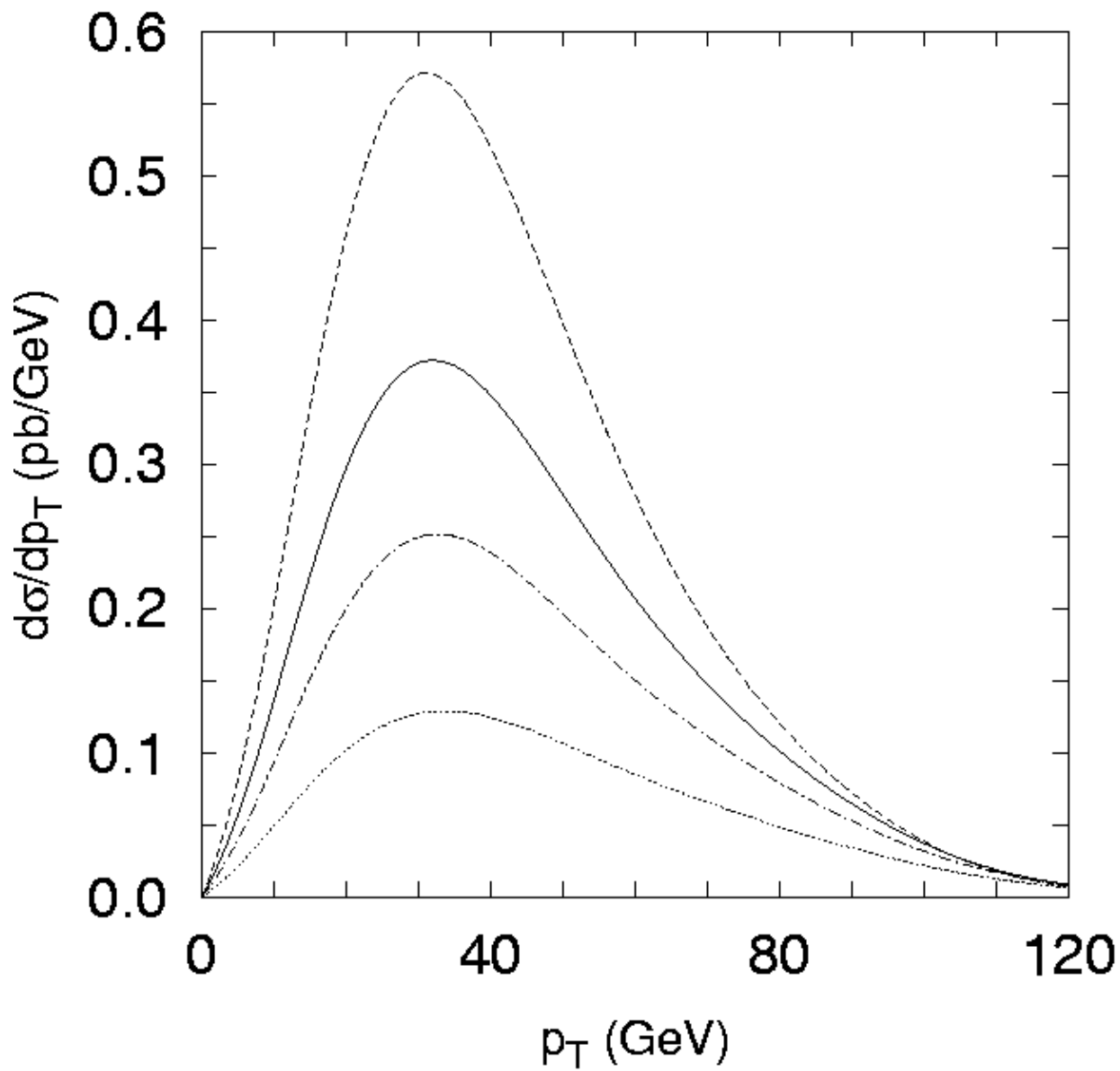}
\end{center}
\caption{Differential cross section $d\sigma/dp_T$ for Higgs boson + one
  jet production at $\sqrt s = 14$ TeV. The kinematical cut 
  $|{\mathbf p}_{{\rm jet}\,T}| > 20$ GeV was applied. 
  All curves are the same as in Figure 2.}
\label{fig7}
\end{figure}

\begin{figure}
\begin{center}
\epsfig{figure=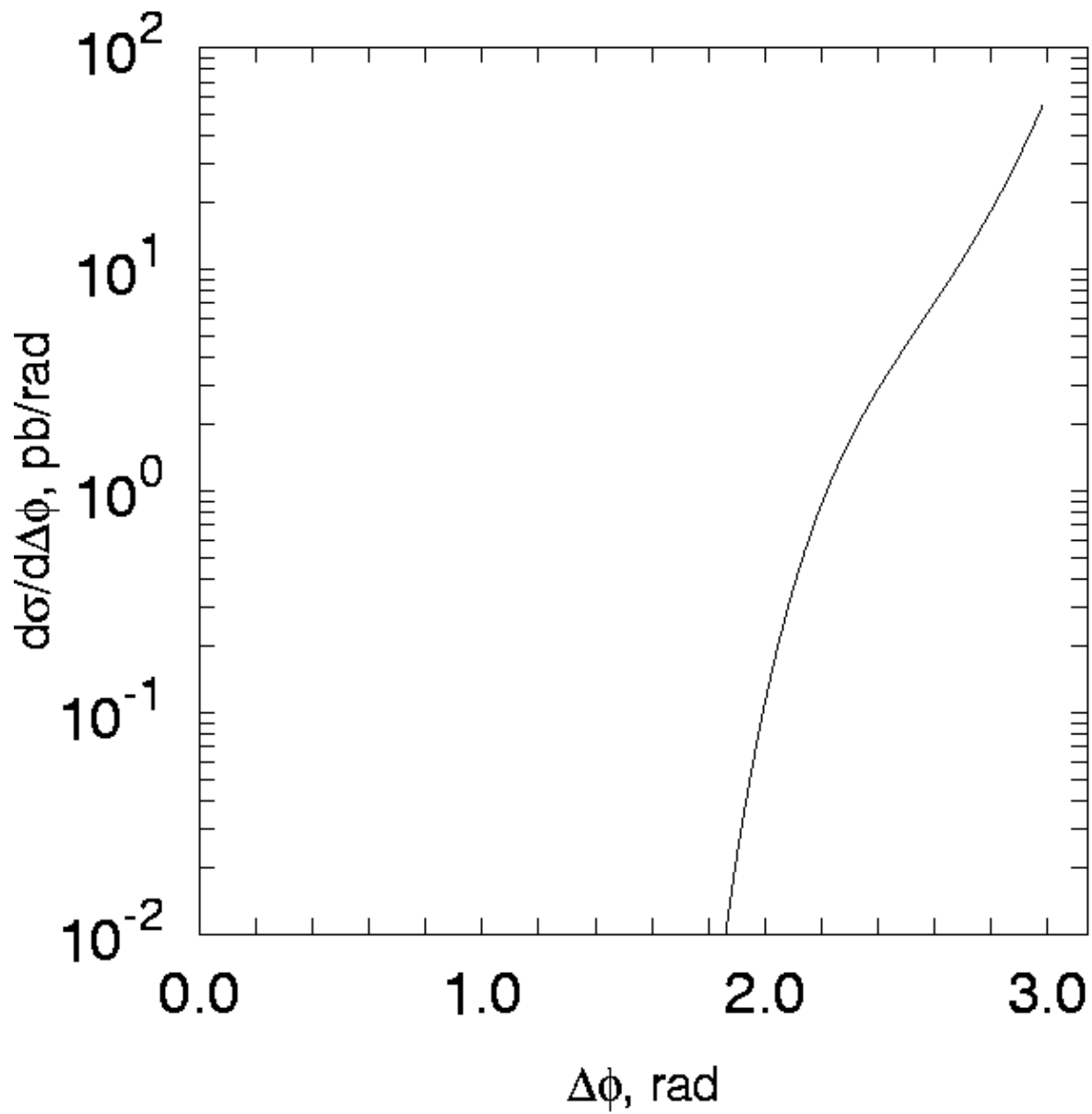}
\end{center}
\caption{The Higg-jet azimutal angle distribution in the 
  Higgs boson + one jet production at $\sqrt s = 14$ TeV. The kinematical cut 
  $|{\mathbf p}_{{\rm jet}\,T}| > 20$ GeV was applied.}
\label{fig8}
\end{figure}

\begin{figure}
\begin{center}
\epsfig{figure=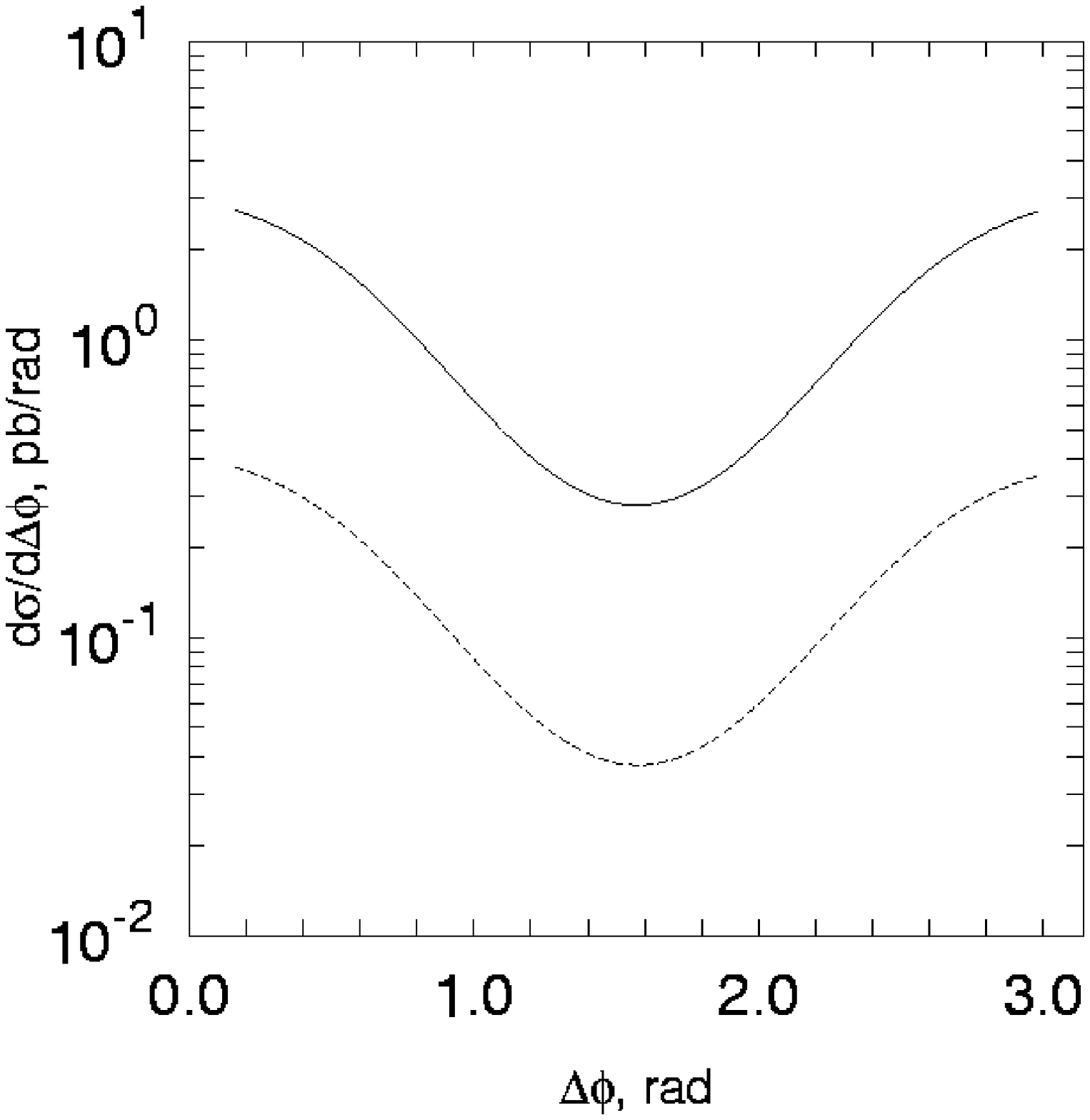}
\end{center}
\caption{The jet-jet azimutal angle distribution in the
  Higgs boson + two jet production at $\sqrt s = 14$ TeV. The kinematical cut 
  $|{\mathbf p}_{{\rm jet}\,T}| > 20$ GeV was applied for both jets.
  Solid and dashed lines correspond to the J2003 set 1 and J2003 set 2 
  unintegrated gluon distributions, respectively.}
\label{fig9}
\end{figure}


\begin{thebibliography}{36}

\bibitem{1} ATLAS Collaboration, Technical Design Report, Vol. 2, CERN/LHCC/99-15, 1999;\\
  CMS Collaboration, Technical Proposal, CERN/LHCC/94-38, 1994.
\bibitem{2} F.~Wilczek, Phys. Rev. Lett. {\bf 39}, 1304 (1977);\\
  H.M.~Georgi, S.L.~Glashow, M.E.~Machacek and D.V.~Nanopoulos, {\it ibid.} Phys. Rev. Lett. {\bf 40}, 692 (1978);\\
  J.R.~Ellis, M.K.~Gaillard, D.V.~Nanopoulos and C.T.~Sachrajda, Phys. Lett. {\bf B83}, 339 (1979);\\
  T.G.~Rizzo, Phys. Rev. {\bf D22}, 178 (1980); {\bf D22}, 1824 (1980).
\bibitem{3} D.~Graudenz, M.~Spira and P.M.~Zervas, Phys. Rev. Lett. {\bf 70}, 1372 (1993);\\
  M.~Spira, A.~Djouadi, D.~Graudenz and P.M.~Zervas, Nucl. Phys. {\bf B453}, 17 (1995).
\bibitem{4} N.~Kauer, T.~Plehn, D.~Rainwater and D.~Zeppenfeld, Phys. Lett. {\bf B503}, 113 (2001);\\
  T.~Plehn, D.~Rainwater and D.~Zeppenfeld, Phys. Rev. {\bf D61}, 093005 (2000);\\
  D.~Rainwater and D.~Zeppenfeld, JHEP {\bf 9712}, 005 (1997).
\bibitem{5} V.N.~Gribov and L.N.~Lipatov, Yad. Fiz. {\bf 15}, 781 (1972);\\
  L.N.~Lipatov, Sov. J. Nucl. Phys. {\bf 20}, 94 (1975);\\
  G.~Altarelly and G.~Parizi, Nucl. Phys. {\bf B126}, 298 (1977);\\
  Y.L.~Dokshitzer, Sov. Phys. JETP {\bf 46}, 641 (1977).
\bibitem{6} E.A.~Kuraev, L.N.~Lipatov and V.S.~Fadin, Sov. Phys. JETP {\bf 44}, 443 (1976);\\
  E.A.~Kuraev, L.N.~Lipatov and V.S.~Fadin, Sov. Phys. JETP {\bf 45}, 199 (1977);\\
  I.I.~Balitsky and L.N.~Lipatov, Sov. J. Nucl. Phys. {\bf 28}, 822 (1978).
\bibitem{7} V.N.~Gribov, E.M.~Levin and M.G.~Ryskin, Phys. Rep. {\bf 100}, 1 (1983).
\bibitem{8} E.M.~Levin, M.G.~Ryskin, Yu.M.~Shabelsky and A.G.~Shuvaev, Sov. J. Nucl. Phys. {\bf 53}, 657 (1991).
\bibitem{9} S.~Catani, M.~Ciafoloni and F.~Hautmann, Nucl. Phys. {\bf B366}, 135 (1991).
\bibitem{10} J.C.~Collins and R.K.~Ellis, Nucl. Phys. {\bf B360}, 3 (1991).
\bibitem{11} M.~Ciafaloni, Nucl. Phys. {\bf B296}, 49 (1988);\\
  S.~Catani, F.~Fiorani and G.~Marchesini, Phys. Lett. {\bf B234}, 339 (1990);\\
  S.~Catani, F.~Fiorani and G.~Marchesini, Nucl. Phys. {\bf B336}, 18 (1990);\\
  G.~Marchesini, Nucl. Phys. {\bf B445}, 49 (1995).
\bibitem{12} J.R.~Forshaw and A. Sabio Vera, Phys. Lett. {\bf B440}, 141 (1998).
\bibitem{13} B.R.~Webber, Phys. Lett. {\bf B444}, 81 (1998).
\bibitem{14} G.P.~Salam, JHEP {\bf 03}, 009 (1999).
\bibitem{15} B.~Andersson {\sl et al.} (Small-$x$ Collaboration), Eur. Phys. J. {\bf C25}, 77 (2002).
\bibitem{16}  V.~Del~Duca, W.~Kilgore, C.~Olear, C.~Schmidt and 
D.~Zeppenfeld, Nucl. Phys. {\bf B616}, 367 (2001);
 Phys. Rev. {\bf D67}, 073003 (2003). 
\bibitem{17} J.R.~Ellis, M.K.~Gaillard and D.V.~Nanopoulos, Nucl. Phys. 
{\bf B106}, 292 (1976);\\
  M.A.~Shifman, A.I.~Vainstein, M.B.~Voloshin and V.I.~Zakharov, Yad. Fiz. {\bf 30}, 1368 (1979).
\bibitem{18} R.V.~Harlander and W.B.~Kilgore, Phys. Rev. Lett. {\bf 88}, 
201801 (2002);\\
  C.~Anastasiou and K.~Melnikov, Nucl. Phys. {\bf B646}, 220 (2002);\\
V.~Ravindran, J.~Smith and W.L.~van~Neerven, Nucl. Phys. {\bf B665}, 325 
(2003).
\bibitem{19} S.~Dawson, Nucl. Phys. {\bf B359}, 283 (1991);\\
  A.~Djouadi, M.~Spira and P.M.~Zervas, Phys. Lett. {\bf B264}, 440 (1991).
\bibitem{20} V.~Del Duca, hep-ph/0312184.
\bibitem{21} J.C.~Collins and D.E.~Soper, Nucl. Phys. {\bf B193}, 381 
(1981); {\it ibid.} {\bf B213}, 545 (1983); {\bf B197}, 446 (1982).
\bibitem{22} J.C.~Collins, D.E.~Soper and G.~Sterman, Nucl. Phys. {\bf 
B250}, 199 (1985).
\bibitem{23} R.K.~Ellis and S.~Veseli, Nucl. Phys. {\bf B511}, 649 
(1998);\\
  R.K.~Ellis, D.A.~Ross and S.~Veseli, {\it ibid.} Nucl. Phys. {\bf B503}, 309 (1997).
\bibitem{24} S.~Catani, E.~D'Emilio and L.~Trentadue, Phys. Lett. {\bf 
B211}, 335 (1988);\\
  R.P.~Kauffmann, Phys. Rev. {\bf D45}, 1512 (1992).
\bibitem{25} D.~de~Florian and M.~Grazzini, Phys. Rev. Lett. {\bf 85}, 
4678 (2000); Nucl. Phys. {\bf B616}, 247 (2001).
\bibitem{26} A.~Gawron and J.~Kwiecinski, Phys. Rev. {\bf D70}, 014003 
(2004).
\bibitem{27} M.G.~Ryskin, A.G.~Shuvaev and Y.M.~Shabelski, Phys. Atom. 
Nucl. {\bf 64}, 120 (2001).
\bibitem{28} H.~Jung, Mod. Phys. Lett. {\bf A19}, 1 (2004).
\bibitem{29} G.~Watt, A.D.~Martin and M.G.~Ryskin, Phys. Rev. {\bf D70}, 
014012 (2004), Erratum: {\it ibid.}  {\bf D70}, 079902 (2004), 
hep-ph/0309096.
\bibitem{30} B.R.~Webber, Nucl. Phys. Proc. Suppl. {\bf C18}, 38 (1991);\\
  G.~Marchesini and B.R.~Webber, Nucl. Phys. {\bf B386}, 215 (1992);\\
  A.~Gawron and J.~Kwiecinski, Acta. Phys. Polon. {\bf B34}, 133 (2003).
\bibitem{31} M.A.~Kimber, A.D.~Martin and M.G.~Ryskin, Phys. Rev. {\bf 
D63}, 114027 (2001).
\bibitem{32} H.~Jung, Comput. Phys. Comm.  {\bf 143}, 100 (2002).
\bibitem{33} F.~Hautmann, Phys. Lett. {\bf B535}, 159 (2002).
\bibitem{34} J.~Andersen {\sl et al.} (Small-$x$ Collaboration), Eur.
Phys. J. {\bf C35}, 67 (2004).
\bibitem{35} G.P.~Lepage, J. Comput. Phys. {\bf 27}, 192 (1978). 
\bibitem{36} A.V.~Lipatov and N.P.~Zotov, hep-ph/0412275, submitted to Eur. Phys. J. C. 
\bibitem{37} S.~Catani, D.~de~Florian and M.~Grazzini, Nucl. Phys. {\bf B596}, 299 (2001);\\ JHEP {\bf 0201}, 015 (2002);\\
  G.~Bozzi, S.~Catani, D.~de~Florian and M.~Grazzini, Phys. Lett. {\bf 
B564}, 65 (2003);\\
S.~Catani, D.~de~Florian, M.~Grazzini and P.~Nason, JHEP {\bf 0307},
028 (2003). 
\bibitem{38} S.P.~Baranov and N.P.~Zotov, Phys. Lett. {\bf B491}, 111 
(2000).
\bibitem{39} T.~Plehn, D.~Rainwater and D.~Zeppenfeld, Phys. Rev. Lett. 
{\bf 88}, 051801 (2002).


\end{thebibliography}
\end{document}